\documentclass[preprint]{revtex4}
\usepackage[dvipdfmx]{graphicx}
\usepackage{ulem}
\usepackage{comment}
\begin{document}
%\preprint{}

\title{Sensitivity of nuclear statistical equilibrium to nuclear uncertainties during stellar core collapse}
%nuclear bulk property, mass data and partition function}
%compression and  full ensemble of heavy nuclei for hot and dense matter}
% repeat the \author. \affiliation  etc. as needed

\author{Shun Furusawa}
\email{shun.furusawa@riken.jp}
\affiliation{Interdisciplinary Theoretical and Mathematical Sciences Program (iTHEMS), RIKEN, Wako, Saitama 351-0198, Japan}
%\author{Hiroki Nagakura}
%\affiliation{TAPIR, Walter Burke Institute for Theoretical Physics, Mailcode 350-17, California %Institute of Technology, Pasadena, CA 91125, USA}
%\author{}
%\thanks{}
%\affiliation{}


\date{\today}

\begin{abstract}
%%{\bf
I have systematically investigated 
the equations of state (EOSs) in nuclear statistical equilibrium
under thermodynamic conditions relevant for core collapse of massive stars 
 by varying the  bulk properties of nuclear matter,
the mass data for neutron-rich nuclei, and the finite-temperature modifications of the nuclear model.
 It is found that the temperature dependence of 
the nuclear free energies 
has a significant impact on the entropy and nuclear composition,
 which affect the dynamics of core-collapse supernovae.
There is a little influence from  the bulk properties and the mass data.
For all models, common nuclei that are likely to contribute to core-deleptonization are those 
near $Z\approx30$ and $N\approx50$.
A model with a semi-empirical expression for internal degrees of freedom,  however,  overestimates
 the number densities of magic nuclei with $N\approx50$ and $82$,
while  a model, in which nuclear shell effects are not considered, underestimates  
the number densities of heavy nuclei, and especially of the magic nuclei.
Other models, which include 
the temperature dependence of shell effects
in the internal degrees of freedom and/or in the nuclear  internal free energy, 
indicate that the difference in population between magic nuclei and non-magic nuclei disappears as the temperature increases. 
The construction of complete statistical EOS will require further theoretical and experimental studies of medium-mass, neutron-rich nuclei with proton numbers 25--45 and neutron numbers 40--85.
\end{abstract}

%\keywords{}

\maketitle

\section{Introduction \label{intro}}
Hot, dense stellar matter appears in  explosive astrophysical phenomena
 such  as core-collapse supernovae and neutron-star mergers,
 but its equation of State (EOS) and weak-interaction rates are not well understood.
The EOS determines the thermodynamic quantities and composition
in terms of nucleons and various nuclei, as functions of temperature, density, and charge fraction.
In the near future, gravitational waves from neutron-star mergers may provide us
 with information about thermodynamic quantities such as pressure at supranuclear density \cite{abott17,shibata17}.
In contrast,
neutrinos from core-collapse supernova may enable investigations of 
the nuclear composition
of  the EOS and the weak interaction rates at subnuclear densities.

Nuclear electron captures and coherent neutrino-nucleus scattering 
during the collapse of a  massive star play important roles in supernova dynamics and neutrino observations.
In particular, the evolution of the charge fraction and entropy in the collapsing core 
is very sensitive to weak interactions  involving  nuclei that appear at densities of  $\rho_B \approx10^{11}\textrm{--}10^{12}$~g/cm$^3$ immediately before the onset of neutrino trapping. %, 
%essentially determine .
As a result, the nuclear weak interactions potentially affect the total mass of a newborn neutron star, the peak neutrino luminosity, and the strength of a shock wave \cite{sullivan16}. 
Recently, some groups have suggested that medium-mass, neutron-rich nuclei may dominate
 the deleptonization process \cite{sullivan16,furusawa17b,raduta16,raduta17,titus18},
although the approach used to determine the nuclear composition or the nuclear model in the EOS  differs from group to group. 

%There are several works comparing
Several investigations have compared different  EOS models \cite{buyukcizmeci13,hempel15, oertel17}
but have not been systematic in their approaches; therefore, it is difficult to obtain quantitative conclusions. %quantitative discussion is difficult.
Many research groups have  already compared modern statistical EOSs that include full ensemble of nuclei, and classical single-nucleus EOSs in that a representative nucleus is used. The latter have been shown to exhibit defects  \cite{burrows84,souza09,gulminelli15,furusawa17b,furusawa17c}; accordingly,
this study focuses only on the former in this work.  
%{\bf
The main inputs for the statistical EOSs are experimental data 
and theoretical estimations for uniform nuclear matter, nuclear masses, and nuclear excitations, which are related to each other. 
%}%\endbf

Bulk properties, such as incompressibility, characterize uniform nuclear matter, 
and their roles in supernovae have been  discussed previously, with the primary focus being the stiffness
 of nuclear matter around nuclear densities. 
It is known that softer bulk properties lead to faster contractions of proto-neutron stars,
 higher neutrino luminosities, and the generation of more acoustic waves,
 resulting in more energetic explosions (see, e.g., Refs. \cite{suwa13,nagakura18a}).
 Most supernova EOSs employ relativistic mean-field calculations  \cite{hempel10,typel10,furusawa11,furusawa13a,furusawa17a,sheng11a,sheng11b,steiner13} or Skrme-type interactions \cite{schneider17}, using fixed parameter sets. 
Recently, some realistic nuclear EOSs  based on a variational method that employs the bare nuclear potentials have also been presented \cite{togashi17,furusawa17d}.

Nuclear masses are another essential ingredient for the statistical EOSs as the nuclear composition
 at low densities and temperatures  is sensitive  to the nuclear binding energies.
 During  core collapse, not only do nuclei close to the stability line  appear but so do extremely heavy and/or neutron-rich nuclei;
%For such nuclei,
we require theoretical mass data for such nuclei.
Blinnikov {\it et al.} \cite{blinnikov11} and the FYSS EOSs \cite{furusawa11,furusawa13a,furusawa17a} utilize the KTUY mass data \cite{koura05},
whereas the 
SHO EOS \cite{sheng11a,sheng11b} adopts the old version of FRDM \cite{moller95}.
In addition, the authors of the HS EOSs \cite{hempel10,steiner13} have prepared various EOSs based on different theoretical mass data; see, e.g., Refs. \cite{geng05, moller95,roca08,lalazissis99}.

At finite temperatures, nuclear excitations must be  included in the EOS. 
They increase the internal degrees of freedom, especially for large and non-magic nuclei, which have nucleons in open-shells.
Ideally, in  the statistics,  we should precisely count all the states with individual excitation energies, for all nuclei; however, 
 experimental data for the nuclear level densities are insufficient. 
In addition, theoretical calculations, such as shell-model calculation  \cite{tsunoda17},   for all states would be quite difficult. 
For now, we employ two approaches to take account of the finite-temperature effects.
One introduces internal partition functions as a function of temperature 
in determining the  number densities of nuclei.
The other approach phenomenologically represents the ground state and all excited states by a hot nucleus
so that the nuclear internal free energy depends on the temperature and is identical  to nuclear mass at zero temperature. 

The former approach allows us to partly  use  the experimental data on level  densities directly
if the temperature is not too high and  the excitations are weak.
Ishizuka {\it et al.} \cite{ishizuka03} and the HS EOS employ a semi-empirical expression for internal partition functions provided by  Fai and Randrup \cite{fai82}, which depends only on the temperature and mass number.
The SRO EOS  \cite{schneider17} utilizes tabulated partition functions provided by Raucher \cite{rauscher97,rauscher00,rauscher03}, which is  based on  a Fermi-gas approach, using experimental data for level densities and the FRDM mass data \cite{moller95} to evaluate the separation energies of nucleons. 
The latter approach employs a temperature-dependent internal free energy, such as in  the SMSM (statistical model for supernova matter) EOS  \cite{botvina04, botvina10,buyukcizmeci14}.
The SMSM EOS is an extension of the SMM (statistical multifragmentation model) EOS \cite{bondorf95},
 which reproduces  fragmentations in low-energy heavy-ion collisions.
The SHO EOS also assumes similar excitation effects for the bulk free energy of the nuclei.
This approach may be valid for reproducing the ensemble of hot nuclei when the temperature is sufficiently high to make the excited states dominant.
%{\bf
Ground-state properties, such as neutron magic numbers and pair energies, are considered to be almost washed out at $T\approx$ 2--3~MeV \cite{brack74, bohr87, sandulescu97,nishimura14}, which are temperatures typical of the late phases of core collapse.
%}%\endbf
The FYSS EOS \cite{furusawa17a,furusawa17d} employs
 temperature 
interpolation between  the internal free energies of cold and hot nuclei.
%
%  is introduced.

In the present work, I investigate the sensitivity of the nuclear composition to several nuclear uncertainties, the bulk properties of nuclear matter, the nuclear mass data,
and the finite-temperature modeling in order to clarify ambiguities in the supernova EOS at subnuclear densities
and to organize the target properties and nuclides  for future theoretical and experimental studies. 
The organization of this paper is as follows: 
In Sec.~\ref{sec:model}, I describe the formulations and inputs for the EOSs.
Comparisons among  EOSs under thermodynamics conditions during core collapse of massive 15M$_{\odot}$ and 25M$_{\odot}$ stars are discussed in Sec.~\ref{sec:res}.
Finally, a summary is given in Sec.~\ref{sec:conc}

\section{Nuclear physics inputs for statistical equations of state \label{sec:model}}
The free-energy density is minimized with respect to the number densities of all particles,
% under
subject to the constrains of baryon and charge conservation,
to obtain the EOS as a function of the volume-averaged values of the baryon number density $n_B$,
  temperature $T$, and charge fraction $Y_p$.
I systematically change three ingredients and
%:
 Table~\ref{tab_model} summarizes the specific cases.
Below, I explain the 14 models---
1B, 1D, 1E, 2FB, 2FD, 2FE, 2HE, 2KE, 2WE, 3FE, 3HE, 3KE, 3WE, and 4FE
---for which the number and abbreviations denote, in order,  
the treatment of finite-temperature effects for nuclear excitations,
the choice of mass data for models 2--4, and 
the parameter set for bulk nuclear matter.

\begin{figure}
\centering 
\includegraphics[width=12cm]{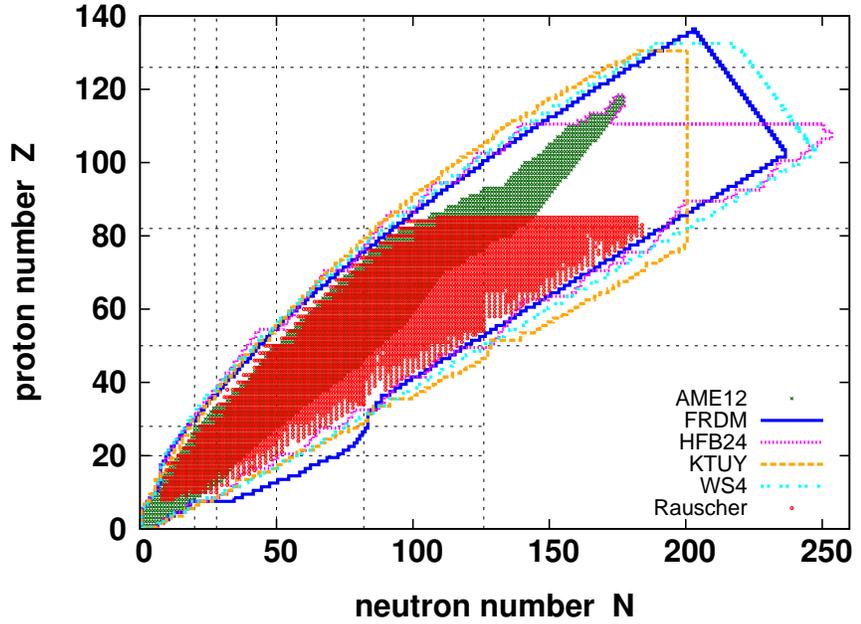}
%\ \\
%\ \\
\caption{Nuclear species for which  experimental nuclear mass data AME12 \cite{audi12} (green crosses) have been adopted.
Theoretical mass data are available in extended regions
 in the (N, Z) plane %by  are 
inside the following contours: FRDM \cite{moller16} (blue solid line),
HFB24 \cite{goriely13} (magenta dotted line), KTUY \cite{koura05} (orange dashed line), and WS4 \cite{wang14} (cyan double-dotted line). 
The red circles indicate nuclei for which the data are obtained from Rauscher's partition function \cite{rauscher03}. 
The black dashed lines denote the neutron- and proton-magic numbers.}
\label{fig_nuclide}
\end{figure}

The constituents of nuclear matter under nuclear statistical equilibrium are the dripped nucleons and all nuclei. 
%{\bf %reply a
The free-energy density can be given by %is expressed as  
\begin{eqnarray}
f  = \xi \left[ n'_p m_p+n'_n m_n +(n'_p+n'_n) \omega \left(n'_p+n'_n,T,\frac{n'_p}{n'_p+n'_n}\right) \right] \nonumber \\
 +  \sum_{AZ}  n_{AZ} \left\{ T \left[ \ln \left(\frac{ n_{AZ}/\kappa}{g_{AZ} (M_{AZ} T/2\pi \hbar ^2 )^{3/2}  }\right)- 1 \right]  + M_{AZ} \right\},
\label{total}
\end{eqnarray}
where $n'_p$ and $n'_n$  are the local
number densities of protons and neutrons, respectively, and $m_p$ and $m_n$, are the corresponding masses
\cite{furusawa18a}. 
The volume factor for free nucleons is defined as $\xi=V'/V$ with the vapor volume $V'$ and the total volume $V$.
The local free energy density of the dripped nucleons
is obtained from a nuclear-matter calculation of $\omega(n_B,T,x)$
 as a function of the local values of  $n_B$, $T$, and the charge fraction $x$,
 as in Furusawa and Mishustin \cite{furusawa18a}. 
The vapor volume can be calculated as $V'=V(1-\sum_{AZ} n_{AZ} A/n_{sAZ})$. 
%}%endbf
The summation covers all nuclides listed in the tabulated data, theoretical mass data or partition functions, as discussed later and shown in Fig.~\ref{fig_nuclide}. 
% are expressed as  $n'_{p}=n_p/\xi$ and  $n'_n=n_n/\xi$.
%, and the local baryon density and charge fraction are expressed as $n_B=n'_p+n'_n$ and $x=n'_p/(n'_p+n'_n)$. 
%, where $n_{sAZ}$ are the nuclear saturation densities.
%, and volume-averaged  number densities of dripped nucleons are expressed as    $n_{p/n}=V'/V n'_{p/n}$.  
%!!! up to Z=200 and N=500??
%
%{\bf 
The quantities $M_{AZ}$, $g_{AZ}$, $n_{AZ}$, and $n_{sAZ}$ are the mass, internal degree of freedom, number density, and saturation density of the nucleus with the mass number $A$ and the proton number $Z$.
%}%endbf
The translational energy of the nuclei is calculated as for an ideal Boltzmann gas, with excluded volume effects represented by  $\kappa=1-n_B/n_0$.

\subsection{Nuclear matter calculation for dripped nucleons and bulk energies of nuclei}
The bulk properties at the nuclear saturation density  $n_0$ for symmetric matter are
characterized by the following parameters: the binding energy at saturation $\omega_0$, the incompressibility $K_0$, 
 symmetry energy at saturation $S_0$, and  symmetry energy slope parameter $L$.
I use the three parameter sets \cite{oyamatsu03,oyamatsu07} listed in Tab~\ref{tab_bulk}.
Parameters B and E share the same value of $K_0$, while parameters D and  E are based on the same value of the saturation slope parameter  $y=-K_0 S_0/(3 n_0 L)$, which gives  the saturation point of asymmetric nuclear matter at about $x=0.5$ as $n_0 +  (1-2x)^2 S_0/y$.
% around $x=0.5$.
The parameters other than $K_0$ and $y$ are optimized to reproduce the masses and radii of stable nuclei by using Thomas--Fermi calculations \cite{oyamatsu03,oyamatsu07}.
The free energy per baryon for uniform nuclear matter, $\omega(n_B,T,x)$, is
% the same in Furusawa and Mishustin \cite{furusawa18a} and 
 based on the  finite-temperature calculation \cite{furusawa18a} of  Oyamatsu and Iida \cite{oyamatsu03,oyamatsu07}
and is given by:
\begin{eqnarray}%
 \label{eq:para1}
 \omega(n_B,T,x) &= &  \omega_{int}+\omega_{kin} , \\
\omega_{int}&=&4 x(1-x) v_s(n_B) /n_B + (1-2 x)^2 v_n(n_B) /n_B ,  \nonumber \\
\omega_{kin}&=& \frac{T}{2 \pi^2 n_B}  \left(\frac{2m_pT}{\hbar^2} \right)^{3/2}   F_{3/2} (\eta_p) +   \frac{T}{2 \pi^2 n_B}  \left( \frac{2m_nT}{\hbar^2} \right)^{3/2}  F_{3/2} (\eta_n)  , \nonumber \\
v_s(n_B) & = &a_1 n_B^2 + \frac{a_2 n_B^3}{1+a_3 n_B}   , \\
v_n(n_B) & = &b_1 n_B^2 + \frac{b_2 n_B^3}{1+b_3 n_B},
\end{eqnarray}
where 
%{\bf 
$\omega_{int}$ and $\omega_{kin}$ are interaction and kinetic energies,  $v_s$ and $v_n$ are the energy densities for symmetric matter and pure-neutron matter, respectively, $b_3=$1.58632~fm$^3$ and 
$\eta_{p/n}=\left[ \mu^0_{p/n} - \partial (\omega_{int} n^0_{p/n})/ \partial n^0_{p/n} \right]/T
$.
%}%endbf
The number densities of protons, $x n_B$, and neutrons, $(1-x) n_B$, correspond to the chemical potentials, $\mu^0_{p/n}$,
and $F_{k}(\eta)$ is defined by $F_{k}(\eta)=\int_0^{\infty} u^{k} \left[ 1+\exp{(u-\eta)} \right]^{-1} du $.
The parameters $a_1$, $a_2$, $a_3$, $b_1$, and $b_2$ for the potential energies
 are fitted to reproduce the bulk parameters in Table~\ref{tab_bulk} \{see Eqs.~(10--14) in Ref.~\cite{furusawa18a}\}.

\begin{figure}
\includegraphics[width=8cm]{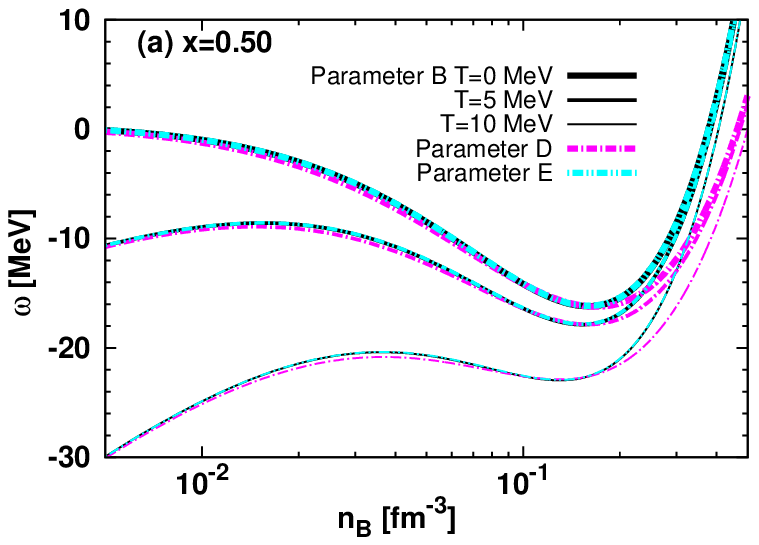}
\includegraphics[width=8cm]{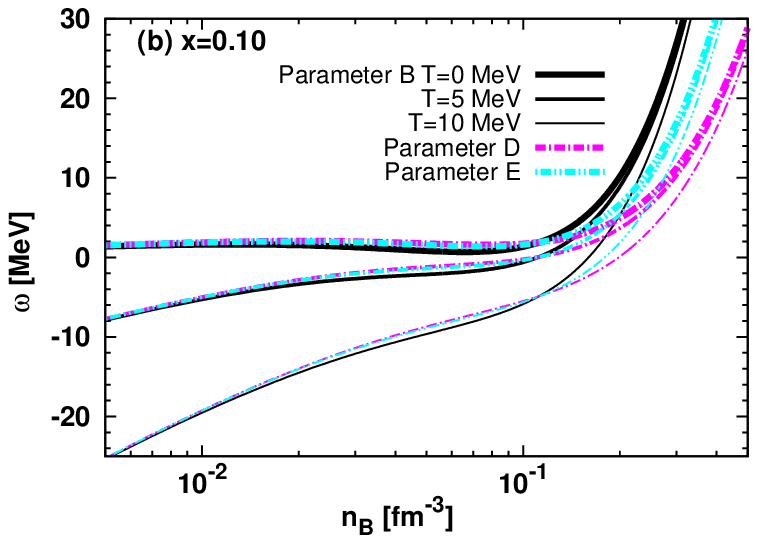}
%\ \\
%\ \\
\caption{Free energy per baryon for symmetric nuclear matter (left panel, $x=0.5$)  and for asymmetric nuclear matter (right panel, $x=0.1$) for parameter sets B (black solid lines), D (magenta dashed-dotted line), and E (cyan double-dotted dashed line)  at $T=$0 MeV (thick lines), 5 MeV (medium lines), and 10 MeV (thin lines). }
\label{fig_blk}
\end{figure}

Figure~\ref{fig_blk} presents the free-energy densities for the three parameter sets.
Differences appear at $n_B\approx 10^{-2}$~fm$^{-3}$.
For symmetric matter  with $x=0.5$, the bulk free energies for parameter sets B and E are almost identical to each other as they have the same value of  $K_0$.
In contrast,  the parameter set D has the soft property of a more gradual
 growth of $\omega$ because of the smaller value of  $K_0$.
%{\bf
For asymmetric matter with $x=0.1$,  
the smaller values of $L$ make the densities higher, at which $\omega$ increases steeply.
%}%endbf
% The smaller absolute value of $y$ leads to a lower saturation density for asymmetric matter,
% and, hence, parameter B provides a lower values of $L$ and the density of  $\omega$-increase.
%The initial characters of model names  in Tab.~\ref{tab_bulk} represent the parameter sets, or models BWL, DWL, and EWL are prepared for the comparison of nuclear bulk properties.

\subsection{Nuclear mass data and model for internal free energy}
The nuclides taken into account in the statistical EOSs are presented in Fig~\ref{fig_nuclide}.
I utilize experimental mass data from AME12  \cite{audi12}
 for those nuclei with data available and  theoretical nuclear mass data from FRDM \cite{moller16}, HFB24 \cite{goriely13}, KTUY \cite{koura05}, or WS4 \cite{wang14}  for the other nuclei.
The middle character---F, H, K, or W---in the names of models 2-4 denotes the specific dataset.
These data also optimize the surface energy parameters for model 2, as discussed later. 
Model 1 includes the nuclides  with FRDM  mass data available, although I do not use the values of nuclear masses. 
%}

%{\it
% nuclear internal free 
%The internal free energy is nuclear mass $M_{AZ}$ at zero temperature in models 1 and 2 and at any temperature in models 3 and 4, as explained later.
%}
%
%{\bf 
In nuclear statistical equilibrium,
the number densities of all nuclei are given by
\begin{eqnarray}
n_{AZ}  =  \kappa \sum_i g_{AZi} \left(\frac{M_{AZ} T}{2\pi \hbar ^2}  \right)^{3/2} {\rm exp} \left(\frac{\mu_{AZ}-M_{AZ}}{T} \right)  , 
%n_{AZ}  =  \kappa \sum_i g_{AZi} \left(\frac{(M_{AZ}+\Delta E_{AZi}) T}{2\pi \hbar ^2}  \right)^{3/2} {\rm exp} \left(\frac{\mu_{AZ}-(M_{AZ}+ \Delta E_{AZi})}{T} \right)  , 
\label{eq:num0} 
\end{eqnarray}
where  the index $i$ runs over the ground state and all excited states, $g_{AZi}$ %and $\Delta E_{AZi}$ 
is the degree of freedom of state $i$,
%and excitation energy of state $i$, 
$\mu_{AZ}= (A-Z) \mu_n + Z \mu_p$, and
 $\mu_n$ and $\mu_p$ are the chemical potentials of neutrons and protons, respectively.
Unfortunately, the experimental data of  the excitation energies or level densities are insufficient at present. 
Instead of including  all excited states explicitly, we often introduce 
 internal degrees of freedom, $g_{AZ}(T)$, and/or internal free energy, $F_{AZ}(T)$, as a function of $T$:
\begin{eqnarray}
n_{AZ}  & \simeq &  \kappa g_{AZ}(T) \left(\frac{M_{AZ}T}{2\pi \hbar ^2}  \right)^{3/2} {\rm exp} \left(\frac{\mu_{AZ}-M_{AZ}}{T} \right)  , \label{nd34} \\ 
          & \simeq & \kappa g_{AZ}^s \left(\frac{M_{AZ}T}{2\pi \hbar ^2}  \right)^{3/2} {\rm exp} \left(\frac{\mu_{AZ}-F_{AZ}(T)}{T} \right).  \label{nd12}
\end{eqnarray}
The former equation can be obtained from the relations $\mu_{AZ}=\partial f/ \partial n_{AZ}$ and Eq.~(\ref{total}).
Models 3 and  4  employ %the temperature-dependent internal degrees of freedom
 $g_{AZ}(T)$ and use cold nuclear masses $M_{AZ}$ that include finite-density effects only, 
as in  Eq.~(\ref{nd34}) and explained later. 
In the models 1 and 2, I substitute 
%temperature-dependent internal free energy 
$F_{AZ}(T)$ %, which contains the effects of nuclear excitations, 
and simplified factors $g_{AZ}^s$
for $M_{AZ}$ and $g_{AZ}(T)$, as in  Eq.~(\ref{nd12}).
For models 1B, 1D and 1E,
I employ a statistical model of the SMSM EOS \cite{botvina04, botvina10,buyukcizmeci14}
 with the nuclear bulk parameters and the formulation of Coulomb energies modified. 
Models 2FB, 2FD, 2FE, 2HE, 2KE,  and 2WE, are based on
the liquid-drop model (LDM) \cite{furusawa18a} with the shell washout effect introduced in the FYSS EOS \cite{furusawa11,furusawa13a,furusawa17a,furusawa17d}.
%Note that there may be other  number-density factors related to thermal wavelength, $(M_{AZ} T)/(2\pi \hbar ^2)$ in this work: $(A m_n T)/(2\pi \hbar ^2)$ as in SMSM EOS, although their differences are negligible. 
%}%endbf

\begin{figure}
\includegraphics[width=11cm]{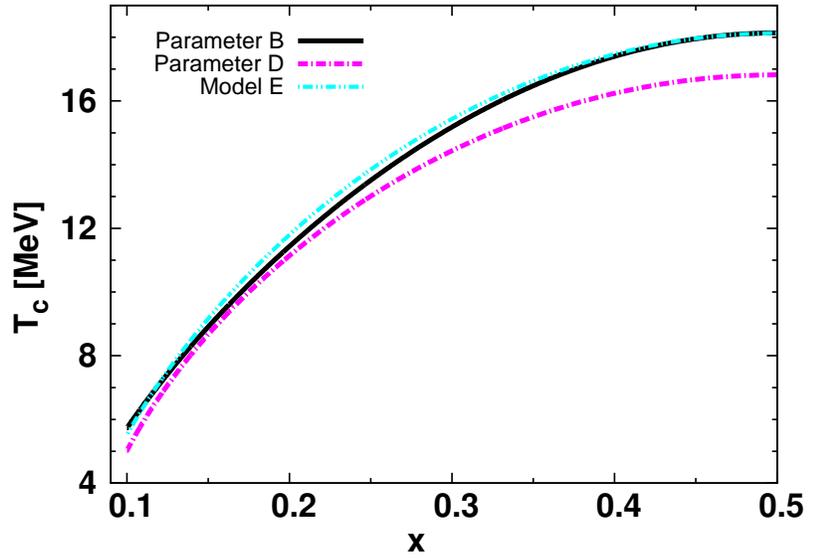}
%\ \\
%\ \\
\caption{Critical temperatures for bulk nuclear matter, above which the bulk pressure has no local minimum,  as a function of charge fraction 
for parameter sets  B (black solid lines), D (magenta dashed-dotted line), and E (cyan double-dotted dashed line).}
\label{fig_ct}
\end{figure}

% 
The internal free energy of model 1 with proton numbers $Z>5$ and neutron numbers  $N>5$
 is composed of nucleon rest masses and bulk, Coulomb, and  surface  energies as follows:
\begin{eqnarray}
F_{AZ} &=& Z m_p + (A-Z) m_n +  F_{AZ}^{B} +  F_{AZ}^{C} + F_{AZ}^{Sf},  \label{eq:mass1} \\
F_{AZ}^{B}&=&  A \left( -\omega_0 - \frac{T^2}{\epsilon} \right) +S_0 \frac{(A-2Z)^2}{A} , \\
F_{AZ}^C&=&\displaystyle{\frac{3}{5}\left(\frac{3}{4 \pi}\right)^{-1/3}  e^2 n_{sAZ}^2 \left(Z/A - n'_p/n_{sAZ} \right)^2 \left({A /n_{sAZ}}\right)^{5/3} D(u_{AZ})} , \label{eq:cl}  \\
F_{AZ}^{Sf}&=& 18 \left( \frac{T^2_{cAZ}-T^2}{T^2_{cAZ}+T^2}\right)^{5/4} A^{2/3}.
\end{eqnarray}
In Eq.~(\ref{eq:cl}),
 $D(u_{AZ})=1-\frac{3}{2}u_{AZ}^{1/3}+\frac{1}{2}u_{AZ}$, where the filling factor $u_{AZ}=(n_e-n'_p)/(Z n_{sAZ} /A -n'_p) $,
 $n_e(=Y_p n_B)$ is the electron number density, and   $e$ is the elementary charge.
In the model, I ignore the iso-spin dependencies of the critical temperature for nuclear vaporization and
 nuclear saturation densities as $T_{cAZ}=18$ MeV and $n_{sAZ}=n_0$.
In model 2, %orm model 1 are the bulk and shell energies, $F_{AZ}^{Sh}$,.
 %differs from model 1in  the bulk 
%we take into account  
I consider nuclear shell effects, $F_{AZ}^{Sh}$, and dependencies of $n_{sAZ}$ on $T$ and on $Z/A$ and employ other formulations for bulk and surface energies.
The internal free energy is given by 
%eon rest masses and bulk, Coulomb,  surface,  and shell energies as follows:
\begin{eqnarray}
F_{AZ} &=& Z m_p + (A-Z) m_n +  F_{AZ}^{B} +  F_{AZ}^{C} + F_{AZ}^{Sf} +   F_{AZ}^{Sh},  \label{eq:mass2} \\
F_{AZ}^{B}&=&  A \{\omega(n_{sAZ},T,Z/A) \},  \label{eq:bulk} \\
F_{AZ}^{Sf}&=&\left(\displaystyle{\frac{36 \pi A^2}{n_{sAZ}^2}} \right)^{1/3}  \sigma_{AZ}(T)
%\left( \frac{\sigma_0 (16 + C_s) \{(T^2_{cAZ}-T^2)/(T^2_{cAZ}+T^2)\}^{5/4}
%} {(1-Z/A)^{-3} + (Z/A)^{-3}  +C_s} 
%\right)
\left(1-\displaystyle{\frac{n'_p+n'_n}{n_{sAZ}}} \right)^2 ,
%\left( \displaystyle{\frac{T^2_{c}(Z/A)-T^2}{T^2_{c}(A/Z)+T^2}} \right)^{5/4}
 \label{eq:sf}  \\ 
 F_{AZ}^{Sh}&=&F_{AZ0}^{Sh} \displaystyle{\frac{\tau_{AZ}}{{\rm sinh}\tau_{AZ}}} . \label{eq:sh} 
\end{eqnarray}
The quantity $F_{AZ}^B$ is based on the same calculations for dripped nucleons as Eq.~(\ref{eq:para1}), and 
%{\bf 
$n_{sAZ}(T)$ is defined as the density at which the free energy, $\omega (n_B,T,Z/A)$, takes its local minimum value around $n_0$.
%}%endbf
The Coulomb energy, $F_{AZ}^{C}$, is calculated by  Eq.~(\ref{eq:cl}).
%In Eq.~(\ref{eq:cl}),
% $D(u_{AZ})=1-\frac{3}{2}u_{AZ}^{1/3}+\frac{1}{2}u_{AZ}$, where the filling factor $u_{AZ}=(n_e-n'_p)/(Z n_{sAZ} /A -n'_p) $,
% $n_e(=Y_p n_B)$ is the electron number density, and   $e$ is the elementary charge.
For $F_{AZ}^{Sf}$ in Eq.~(\ref{eq:sf}), $\sigma_{AZ}(T) =\sigma_0 [(T^2_{cAZ}-T^2)/(T^2_{cAZ}+T^2)]^{5/4} (16 + C_s) /[(1-Z/A)^{-3} + (Z/A)^{-3}  +C_s ] $ and
 $T_{cAZ}$, is defined by using  the bulk pressure, $P_{bulk}=n_B^2 \partial \omega(n_B,T,x=Z/A)/\partial n_B$.
This is  the temperature at which 
$(\partial P_{bulk}/\partial n_B)|_{x}=0$ and $(\partial^2 P_{bulk}/\partial n_B^2)|_{x}=0$  simultaneously and  is shown in Fig.~\ref{fig_ct} for each bulk parameter set.
The last factor in Eq.~(\ref{eq:sh}) expresses the washout effects, where $\tau_{AZ} =2 \pi^2 T/(41 A^{-1/3})$ \cite{furusawa13a,furusawa17a}.
%We refer the reader to refs. \cite{furusawa13a,furusawa17a}  for
% information about $n_{sAZ}$ and $E^{Sh}_{AZ}$ and to ref. \cite{furusawa18a} for 
%the details about  the LDM with washout. %other terms. 
%

The shell energies of the ground states are set equal to the mass data, $M_{AZ}^{data}$,  minus the 
LDM mass-energy in a zero-density medium, i.e., $ F_{AZ0}^{Sh}=M_{AZ}^{data} - [ Z m_p + (A-Z) m_n +  F_{AZ}^{B} +   F_{AZ}^{C} + F_{AZ}^{Sf} ]_{n_e=0,n'_p=0,n'_n=0,T=0}$. 
This assumption allows the internal free energy of Eq.~(\ref{eq:mass2}) to  reproduce  exactly the mass data at low temperatures and densities.
%{\bf 
Here, I assume  the shell energy to be positive to avoid  negative entropy production.
Nuclear excitations usually increase the internal degrees of freedom,
which corresponds to a free-energy reduction associated with a temperature increase and to positive entropy production.
This assumption for the sign of shell
correction, however, is not typical and, in some references \cite{egidy05}, the shell energies of magic nuclei take negative values.
%}%\endbf
The surface tension $\sigma_0$ and isospin-dependence parameter $C_s$ are optimized to minimize the total deviation of the LDM mass-energy (internal free energy in zero-temperature limit) per baryon from the mass data or the sum of the shell energies per baryon, $\sum_{Z>5,n>5}F^{Sh}_{AZ0}/A$.
The resulting values of these quantities are listed in Table~\ref{tab_model}.

\begin{figure}
\includegraphics[width=8cm]{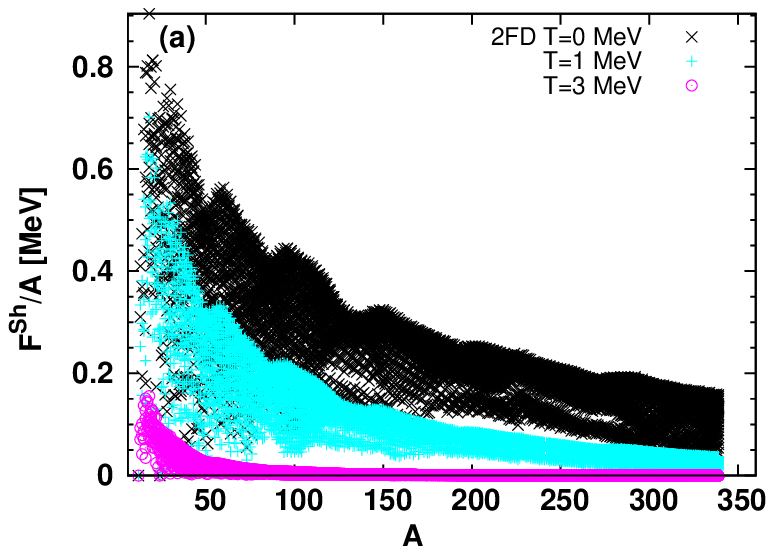}
\includegraphics[width=8cm]{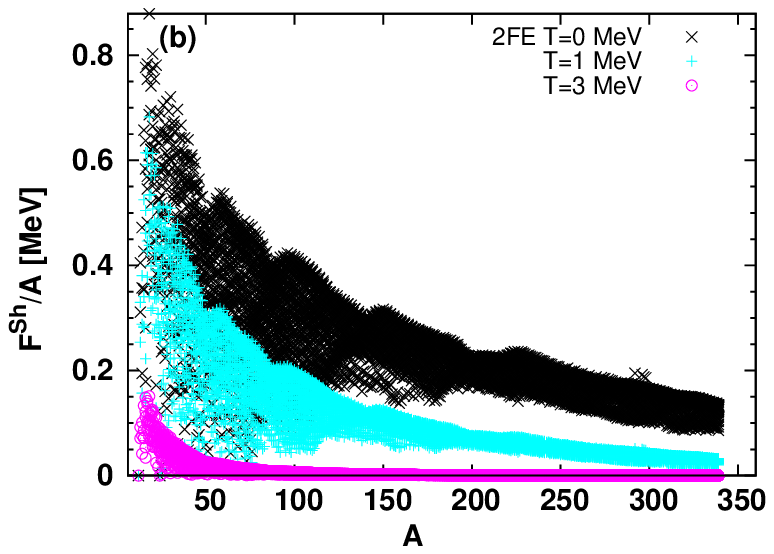}
\includegraphics[width=8cm]{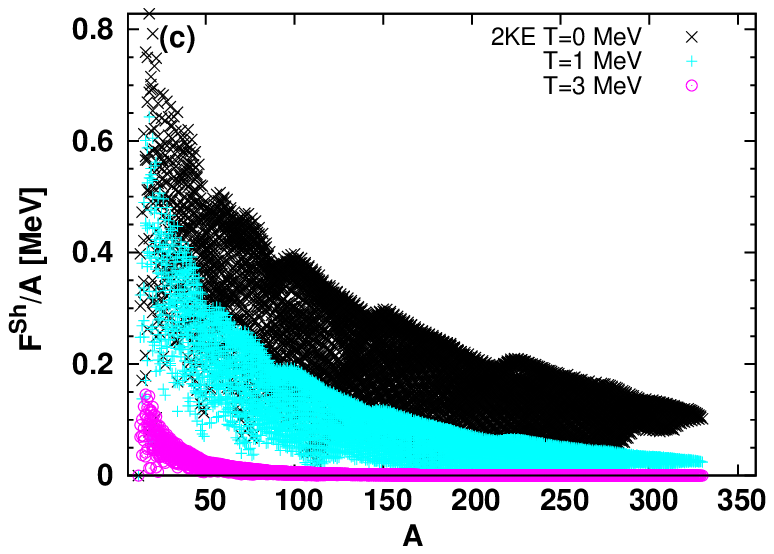}
\includegraphics[width=8cm]{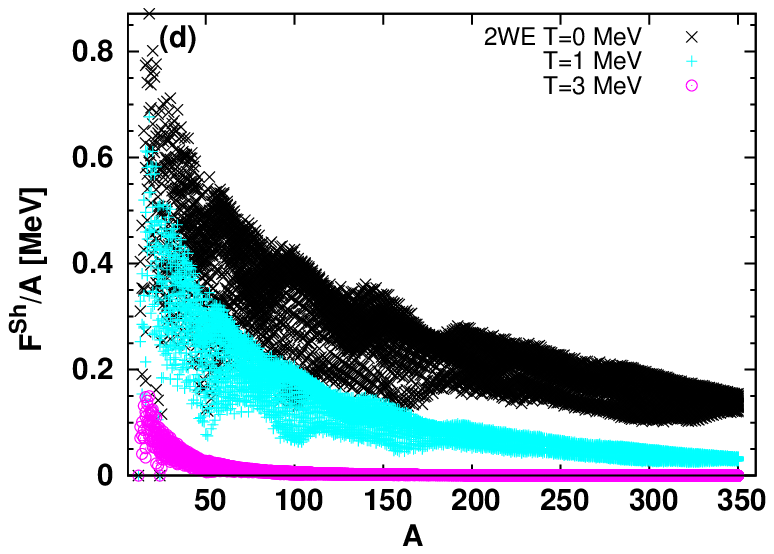}%\ \\
%\ \\
\caption{Shell energies per baryon for models 2FD (top-left panel),  2FE (top-right panel), 2KE (bottom-left panel), and  2WE (bottom-right panel)  at $T=$0 MeV (black crosses),
 1 MeV (cyan pluses), and 3 MeV (magenta circles). }
\label{figshll}
\end{figure}

Figure~\ref{figshll} illustrates the temperature dependence of the shell energies for models 2FD, 2FE, 2KE, and 2WE. The combination of mass data and bulk parameters leads to a set of shell energies and surface tension parameters. At $T=3$ MeV, the nuclear shell effects are significantly reduced, especially for nuclei with large mass numbers.
%{\bf
Note that this formulation is still rough and may overestimate the shell damping  as discussed later.
% In actual, some double magic nuclei such as $^{40}$Ca may retain the shell effects somewhat even over $T=$8 MeV \cite{egidy05}, at which they are almost diminished in model 2.
%}%endbf

%{\bf
In models 3 and 4,
 the temperature dependence  is encapsulated in internal degrees of freedom $g_{AZ}(T)$
as in  Eq.~(\ref{nd34}).
%}%endbf
%arature-dependence in the internal free energies, 
%which mean that   
%and cold nuclear masses are utilized as  $F_{AZ}=M_{AZ}$.
%, as already noted.
To evaluate the nuclear masses, I take into account only Coulomb screening effects: $\Delta E^C_{AZ}=F_{AZ}^C(n_e,n'_p)-F_{AZ}^C(0,0)$ from Eq.~(\ref{eq:cl}),
with $M_{AZ}=M_{AZ}^{data} +\Delta E^C_{AZ}$. 
In models 3 and 4,  I set $n_{sAZ}=n_0$ for all nuclei.

\subsection{Internal degrees of freedom}
In model 3, I use a semi-empirical function for internal degrees of freedom
 that depends only on the temperature and  mass number \cite{fai82}. 
It is utilized in the HS EOSs \cite{hempel10,steiner13}
 and can be given by 
\begin{equation}
\label{eq:ex}
g_{A}(T)=g_{A}^0 +\frac{c_1}{A^{5/3}}\int_0^{16.2 A} dE e^{-E/T}\exp\left(\sqrt{2 a(A) E}\right) , \label{eq:fai} 
\end{equation}
where $a(A)=(A/8)(1-c_2 A^{-1/3})$MeV$^{-1}$, $c_1=0.2$MeV$^{-1}$, $c_2=0.8$,
and $g_{A}^0=1$ for even nuclei and $g_{A}^0=3$ for odd nuclei, and
the upper bound of the integral is set to be a typical bulk energy, $16.2 A$, as in the HS EOS.
%{\bf 
There are several options for the upper limit according to
each model of energy integral: infinity; nuclear binding energies; the smaller of
neutron and proton separation energies with in-medium modifications \cite{gulminelli15,fowler78}.
The choice of the upper bound, however,  does not significantly change the results.
%}%\endbf
%
Model 4FE employs Raucher's tabulated data for the internal partition functions $g_{AZ}(T)$ for each nucleus \cite{rauscher03}, which are used in the SRO EOS  \cite{schneider17}.
The nuclei listed in these data are limited, as shown in Fig~\ref{fig_nuclide}; 
therefore, I assume no excitations for the nuclei,
for which partition function data are unavailable, even if the mass data are available.
%
%The internal degree of freedom for other LDM is defined as $g_{AZ}(T)= (g_{AZ}^0 -1) \tau_i/{\rm sinh}\tau_i  +1$.

%
In  models 1 and 2, 
I consider the excited-state contributions to the quantities
 of $F_{AZ}^{B}$,  $F_{AZ}^{Sf}$, and  $F_{AZ}^{Sh}$  in Eqs.~(\ref{eq:mass1}) and (\ref{eq:mass2}).
The values of $g_{AZ}^s$ in model 1 are set to be unity as in the original EOS. 
%}
In model 2, the ground-state spin factors, $g_{AZ}^0$, are taken from Ref. \cite{rauscher00} for those nuclei for which data are available and are set to $g_{AZ}^0=1$ for other even nuclei and $g_{AZ}^0=3$ for other odd nuclei.
The internal degrees of freedom %for other LDM 
%is defined as $g_{AZ}(T)= (g_{AZ}^0 -1) \tau_i/{\rm sinh}\tau_i  +1$.
%They 
are assumed to be washed out according to $g_{AZ}^s= (g_{AZ}^0 -1) \tau_{AZ}/{\rm sinh}\tau_{AZ}  +1$, approaching unity with
the reduction in the shell energy because of increase in the excited states \cite{furusawa17a}.

For comparison with the internal degrees of freedom for models 3 and 4, I define effective internal degrees of freedom for  models 1 and 2 in refer to Hempel \cite{hempel10d}:
 \begin{eqnarray}
 g^*_{AZ}(T) =g_{AZ}^s {\rm exp} \left(\frac{F^0_{AZ}-F^*_{AZ}}{T} \right)  . \label{eqeff}
\end{eqnarray}  
Here, $F^*_{AZ}$ is the internal free energy in the zero-density-medium limit,
 but at finite temperature, $F_{AZ}(T)=F_{AZ}(T,n_e=0,n'_p=0,n'_n=0)$.
The internal free energy in vacuum limit ($T=0$, $n_e=0$, $n'_p=0$, and, $n'_n=0$),  $F^0_{AZ}$, is identified with $M_{AZ}^{data}$ in model 2.
%} %endef
%Note that not only do the washouts of $F_0^{sh}$ and $g_{AZ}(T)$affect $g^*_{AZ}$
%but so do the reductions in  bulk and surface energies.
This formulation leads to the expression for the number density of nuclei in the zero-density-medium limit:
 \begin{eqnarray}
n_{AZ}(T) =  g^*_{AZ}(T) \left(\frac{F_{AZ}^{0} T}{2\pi \hbar ^2}  \right)^{3/2} {\rm exp} \left(\frac{\mu_{AZ}-F_{AZ}^{0}}{T} \right)  ,
\end{eqnarray} 
which is similar to Eq.~(\ref{nd34}) for models 3 and 4.
The internal degrees of freedom  for these four approaches are presented in Fig.~\ref{figlvl}.
Model 1 generally agrees with Rasucher's data (model 4)  at $T=1$~MeV and
tends to exhibit fewer internal degrees of freedom than model 2 at any temperature.
For model 2, nuclei with small mass numbers are more likely to be excited, 
as compared with the other models,
as their shell and surface energies per baryon are large relative to nuclei with large mass numbers.
The semi-empirical internal degrees of freedom for model 3 are clearly fewer  than those given by 
the other three models.

For all models, light clusters with  $Z \leq5$ or  $N \leq5$ are calculated as ideal Boltzmann gases using the  excluded volume, $\kappa$, 
  Coulomb screening effects,  $F_{AZ}=M_{AZ}=M_{AZ}^{data} +\Delta E^C_{AZ}$, and no excitation, $g_{AZ}=g_0$ as in Ref. \cite{furusawa18a}. 

\section{nuclear statistical equilibrium in collapsing cores \label{sec:res}}
\begin{figure}
\includegraphics[width=12cm]{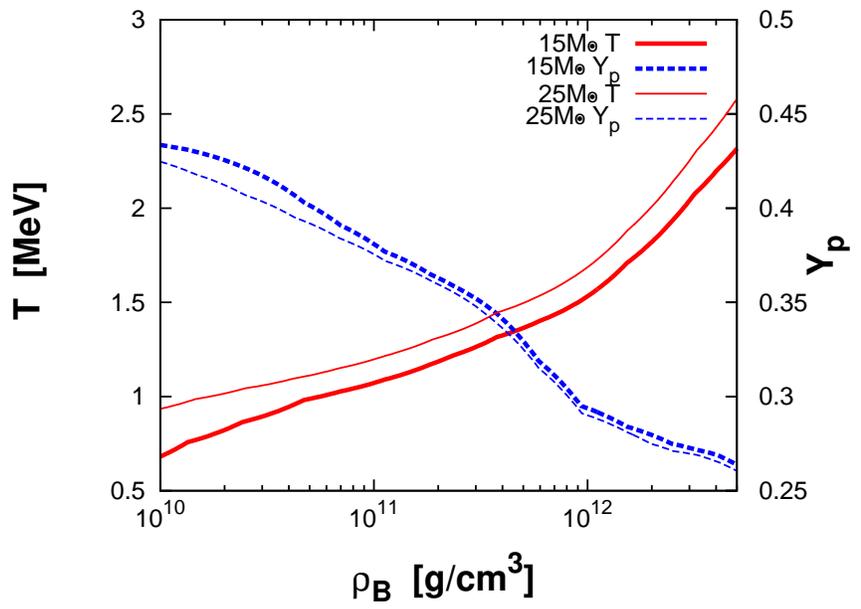}
%\ \\
%\ \\
\caption{
Temperatures (red solid lines) and charge fractions (blue
dashed lines) as a function of density at the centers of the
collapsing cores of supernova progenitors of 15M$_\odot$ (thick lines) and 25M$_\odot$ (thin lines)  \cite{juodagalvis10}.}
\label{fig_pro}
\end{figure}

I next compare some models listed in Table~\ref{tab_model}
 along core-collapse trajectories for stars of 15M$_\odot$ and 25M$_\odot$ \cite{juodagalvis10}.
Figure~\ref{fig_pro} presents the temperature and  charge fraction as a function of density for  these supernova progenitors. 
The progenitor with 25M$_{\odot}$ exhibits higher temperature and lower charge fractions for the same density.
Note that these quantities depend on the EOS and the results of core-deleptonization in actual core-collapse simulations, although I assume the same thermodynamic conditions for all EOSs in this comparison. 
I focus mainly on the nuclear composition of heavy nuclei at $\rho_B \approx10^{11-12}$~g/cm$^3$,
because weak interactions involving heavy nuclei play fundamental roles in core-deleptonization immediately before neutrino sphere formation,
and the impacts of nucleons and light clusters are not dominant.
%the mass fractions of proton and light clusters are  small and dripped neutrons are not related in the deleptonization before the core-bounce.

\begin{figure}
\includegraphics[width=8cm]{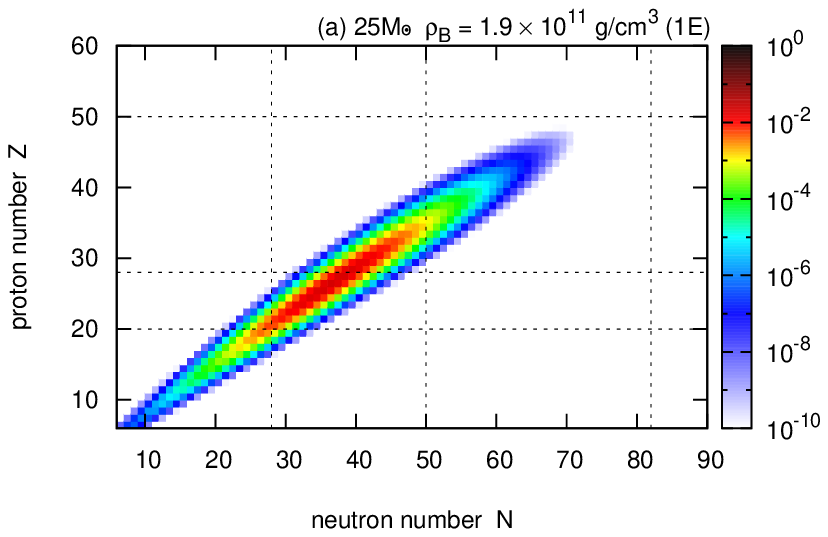}
\includegraphics[width=8cm]{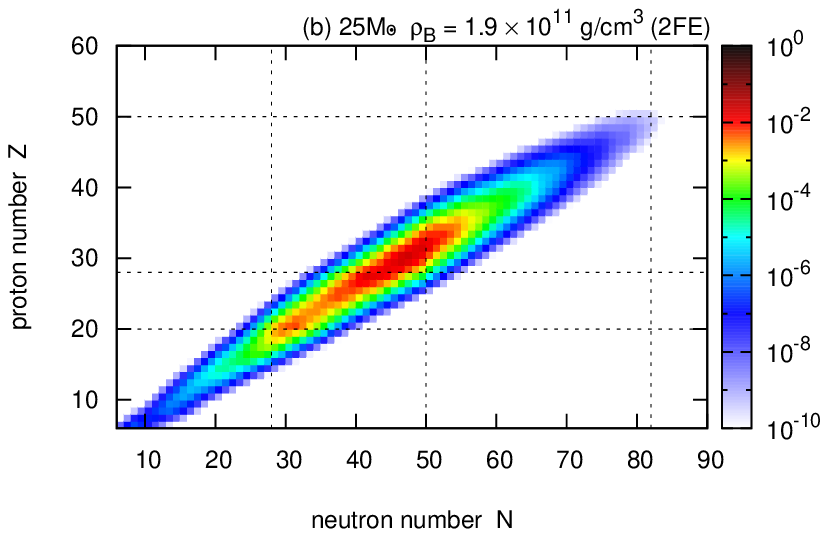} \\
\includegraphics[width=8cm]{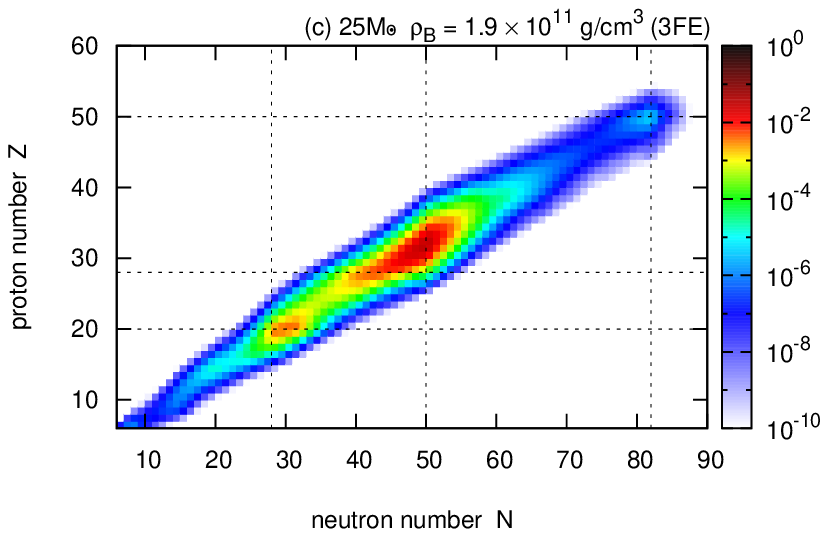}
\includegraphics[width=8cm]{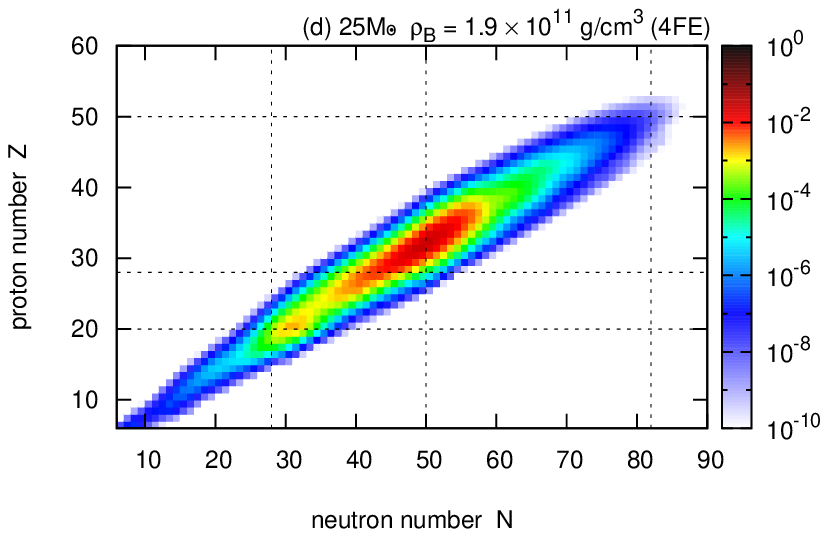}%\\
%\includegraphics[width=8cm]{fig7e.eps}
%\includegraphics[width=8cm]{fig7f.eps}
%\ \\
%\ \\
\caption{
Mass fractions $X_{AZ}$ in the $(N, Z)$ plane  at $(\rho_B, T, Y_p)=(1.9 \times 10^{11}$ g/cm$^3$, 1.3 MeV, 0.36)   for the collapsing core of a 25M$_\odot$
supernova progenitor for models  1E (top-left panel), 2FE (top-right panel), 3FE (bottom-left panel), and  4FE (bottom-right panel). }
\label{fig_dis25}
\end{figure}

\begin{figure}
\includegraphics[width=8cm]{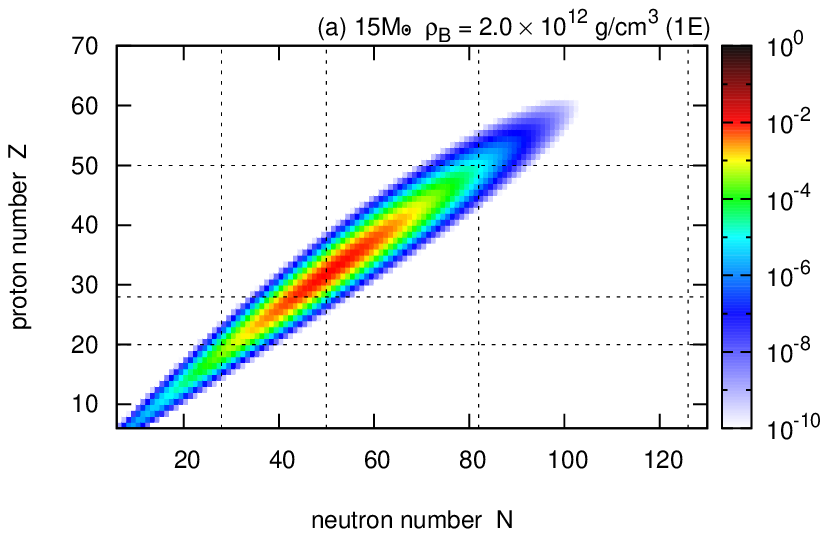}
\includegraphics[width=8cm]{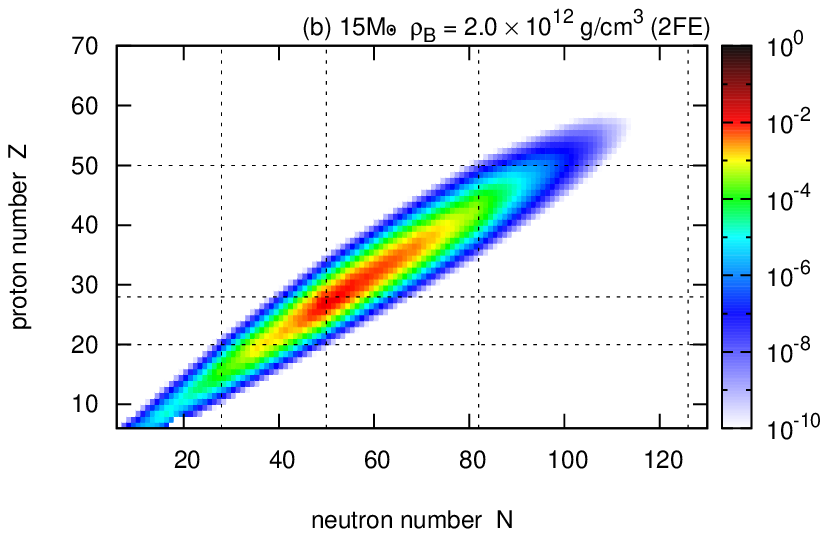} \\
\includegraphics[width=8cm]{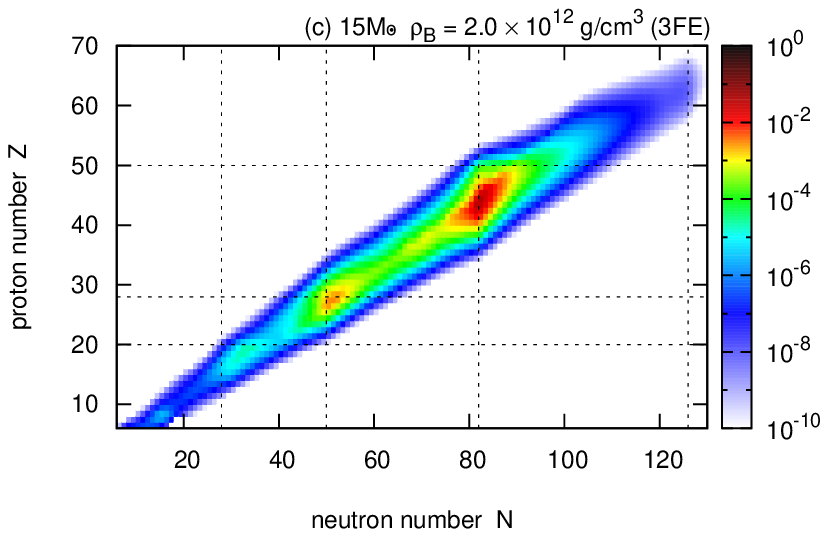}
\includegraphics[width=8cm]{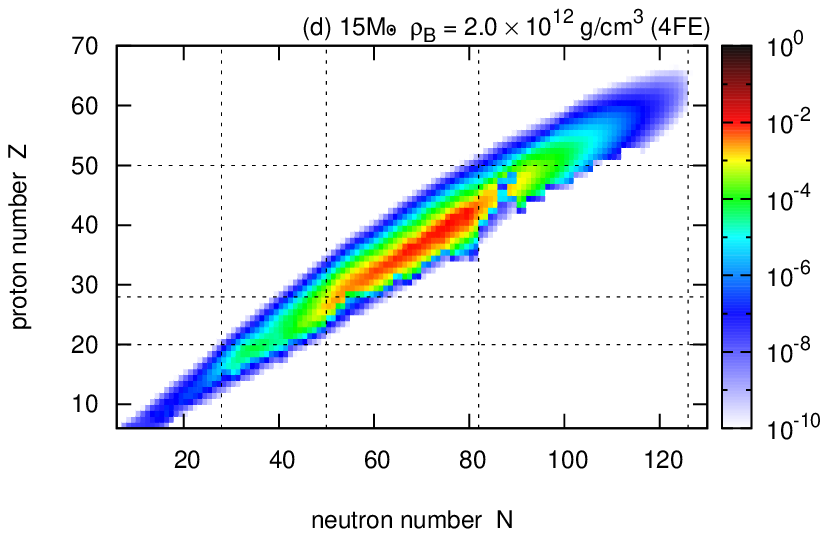}%\\
%\includegraphics[width=8cm]{fig8e.eps}
%\includegraphics[width=8cm]{fig8f.eps}
%\ \\
%\ \\
\caption{
Mass fractions $X_{AZ}$ in the $(N, Z)$ plane  at $(\rho_B, T, Y_p)=(2.0 \times 10^{12}$ g/cm$^3$, 1.8 MeV, 0.28) for the collapsing core of a 15M$_\odot$
 supernova progenitor for models 1E (top-left panel), 2FE (top-right panel), 3FE (bottom-left panel), and  4FE (bottom-right panel). }
\label{fig_dis15}
\end{figure}

Figures~\ref{fig_dis25} and \ref{fig_dis15} display 
the mass fractions of all heavy nuclei, $X_{AZ}=A n_{AZ}/n_B$,
 for models 1E, 2FE, 3FE, and 4FE under thermodynamical conditions 
at $(\rho_B, T, Y_p)=(1.9 \times 10^{11}$ g/cm$^3$, 1.3 MeV, 0.36)  for the 25M$_\odot$-star and
$(2.0 \times 10^{12}$ g/cm$^3$, 1.8 MeV, 0.28) for the 15M$_\odot$-star.
Model 1E shows smoother mass distribution than the other models,
because of the lack of nuclear shell effects.
Other models generally have a monomodal or bimodal structure around neutron magic numbers ($N=$28, 50, and 82).
%}
% exhibits smooth 
 Model 3FE assumes the semi-empirical function for internal degrees of freedom and cold nuclear mass and, as a result, 
overestimates the mass fractions of these magic nuclei. % with $N\approx$~28, 50, and 82.
In model 2FE, the shell energies $F_{AZ}^{Sh}$ are washed out,
 and, consequently, non-magic nuclei are also abundant.
In model 4FE, Rauscher's data for the internal degrees of freedom take into account the shell effects; therefore, non-magic nuclei are more easily  excited and have larger values of $g_{AZ}(T)$ than the magic nuclei.
Therefore,  models 2FE and 4FE produce  wider and smoother distributions
 of  nuclei with large abundances $X_{AZ}$ in the  $(N,Z)$ plane than model 3FE.
Unfortunately, model 4FE lacks the partition function data for some important nuclei close to the abundance peaks, e.g., 
$(N,Z)=(85, 28)$ and (75, 35), as displayed in Fig.~\ref{fig_nuclide}.

\begin{figure}
\includegraphics[width=8cm]{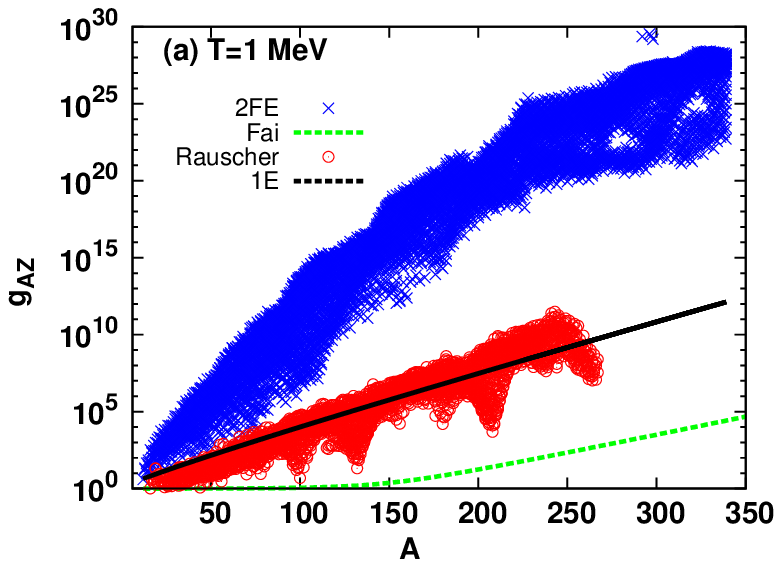}
\includegraphics[width=8cm]{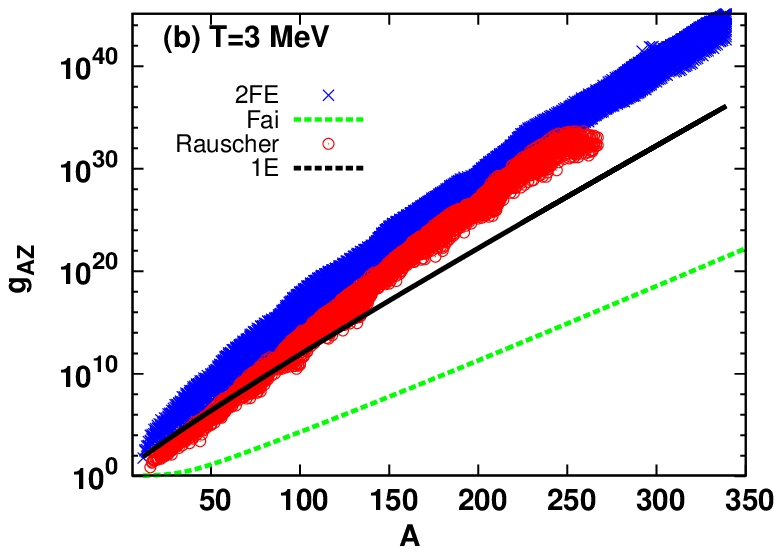}
\includegraphics[width=8cm]{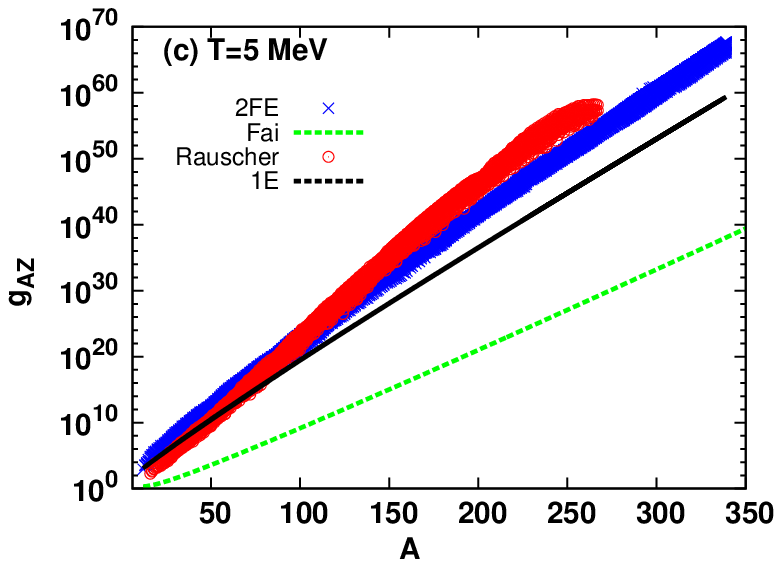}
\includegraphics[width=8cm]{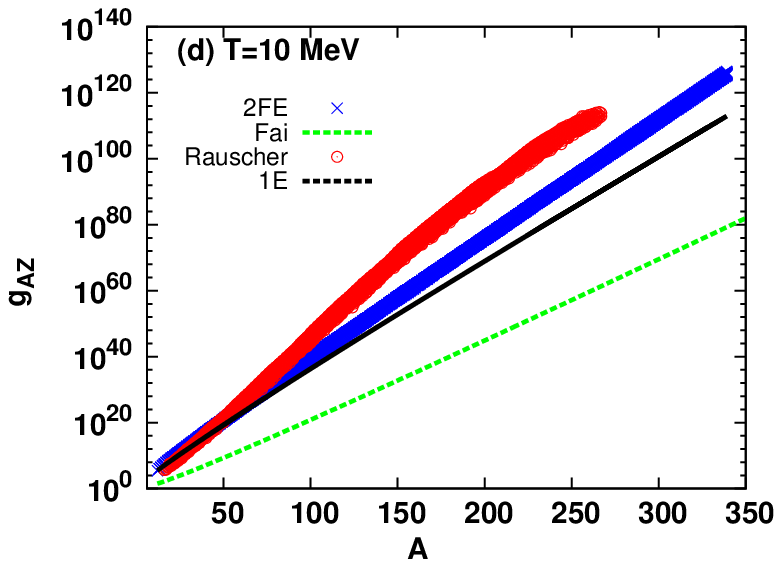}
%\ \\
\caption{Effective internal degrees of freedom for all nuclei for models 1E (black dashed lines) and  2FE (blue crosses), as calculated from Eq.~(\ref{eqeff}); semi-empirical expression for internal degrees of freedom as functions of mass number, as given by Eq.~(\ref{eq:fai}) and provided by Fai and Randrup \cite{fai82} (green dashed lines); and internal partition functions in tabular form of Raucher \cite{rauscher03} (red circle dots) at $T=$1 MeV (top-left panel), 3 MeV (top-right  panel), 5 MeV (bottom-left panel) and 10 MeV (bottom-right panel). }
\label{figlvl}
\end{figure}

\begin{figure}
\includegraphics[width=8cm]{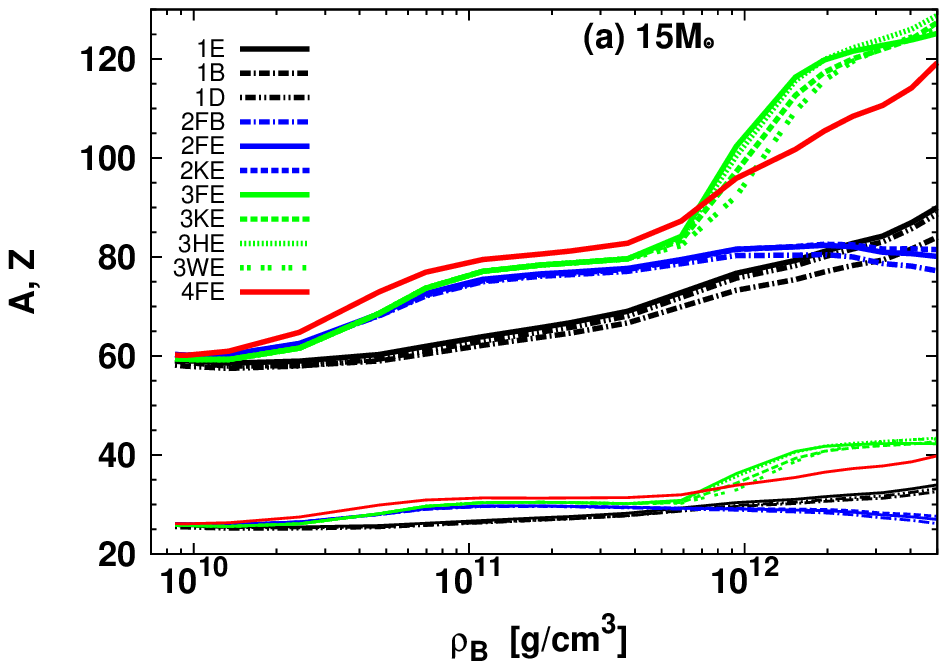}
\includegraphics[width=8cm]{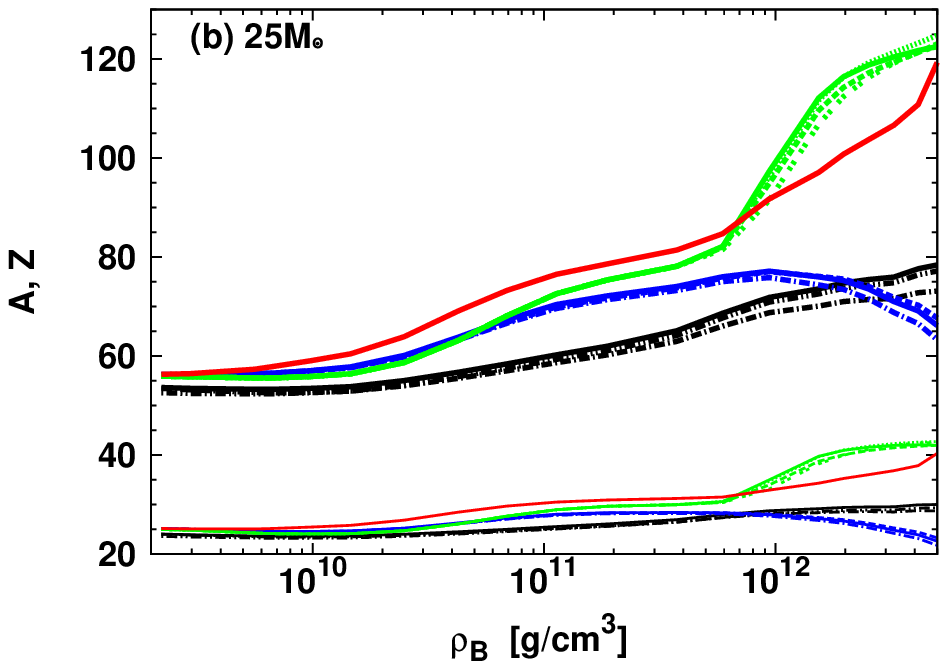}
%\ \\
%\ \\
\caption{Average mass numbers (thick lines) and proton numbers (thin lines) at the centers of the collapsing cores of supernova progenitors of 15M$_\odot$ (left panel) and 25M$_\odot$ (right panel) for models  
1B (black dashed-dotted lines), 1D (black double-dotted dashed lines), 1E (black solid lines),
 2FB (blue dashed-dotted lines),  2FE (blue solid lines),  2KE (blue dashed lines), 
 3FE (green solid lines), 3KE (green dashed lines), 3HE (green dotted lines), 3WE (green double-dotted lines), and  4FE (red solid lines). }
\label{fig_maz}
\end{figure}

The average mass numbers and proton numbers of heavy nuclei are 
shown in Fig.~\ref{fig_maz} for almost all the models.
The impact on these quantities of the mass data and bulk properties is not large,
 while the choice of partition function alters them greatly.
%n contrast, 
%The partition functions change the nuclear compositions considerably as shown in Figs.~\ref{fig_dis25} and \ref{fig_dis15}, and the average mass numbers.
% 
The average mass numbers for models 1B, 1D, and 1E are smaller and grow more gradually  than the other models  owing to the lack of shell effects. 
%}
For models 2FB, 2FE, and 2KE, nuclei with smaller mass numbers 
are more highly excited than those in the other models,
 as discussed above and shown in Fig.~\ref{figlvl}.
Therefore, they tend to produce smaller average mass numbers than do models 3 and 4.
As discussed above and displayed in Figs.~\ref{fig_dis25} and \ref{fig_dis15}, 
nuclear shell effects remain at high temperatures in model 3FE.
Hence, only that model leads to a step-wise growth
 in the average mass numbers and proton numbers, 
owing to the neutron-magic numbers: at about  $\rho_B=10^{12}$~g/cm$^3$, 
the dominant constituents of nuclear matter abruptly change nuclei with $N\approx50$ to those with  $N\approx82$.
% in nuclear chart.
%Model 4FE  shows the larger mass numbers than model 3FE.
%The increase of $g_{AZ}$ or $g^*_{AZ}$  as a function of mass numbers  mainly determine the
% average mass numbers. 

% 
The bulk parameter does not affect significantly the results among models 2.
This is due to  the internal free energies being almost equal to the mass data at low temperatures owing to the introduction of shell energies 
 and to the parameters for nuclear bulk energies other than $K_0$  and $y$ and for  surface tensions also being optimized to reproduce the similar nuclear properties, such as nuclear masses and radii.
The differences arising from the bulk properties are more visible among models 1.
 The larger values of $S_0$ and $L$  reduce 
the binding energies and the average mass numbers 
in models 1B and 2FB compared with  1D, 1E, and  2FE. 
The choice of theoretical mass data does not significantly influence the results either
in models 2, as nuclei for which the experimental mass data exist are dominant.
 Differences from the theoretical mass data appear only for neutron-rich nuclei at high temperatures, where the shell effects derived from the mass data no longer survive.
Models 3FE, 3HE, 3KE, and 3WE display larger differences  due to the mass data than models 2KE and 2FE,
because they do not take into account the shell damping.
% reply 4

\begin{figure}
\includegraphics[width=8cm]{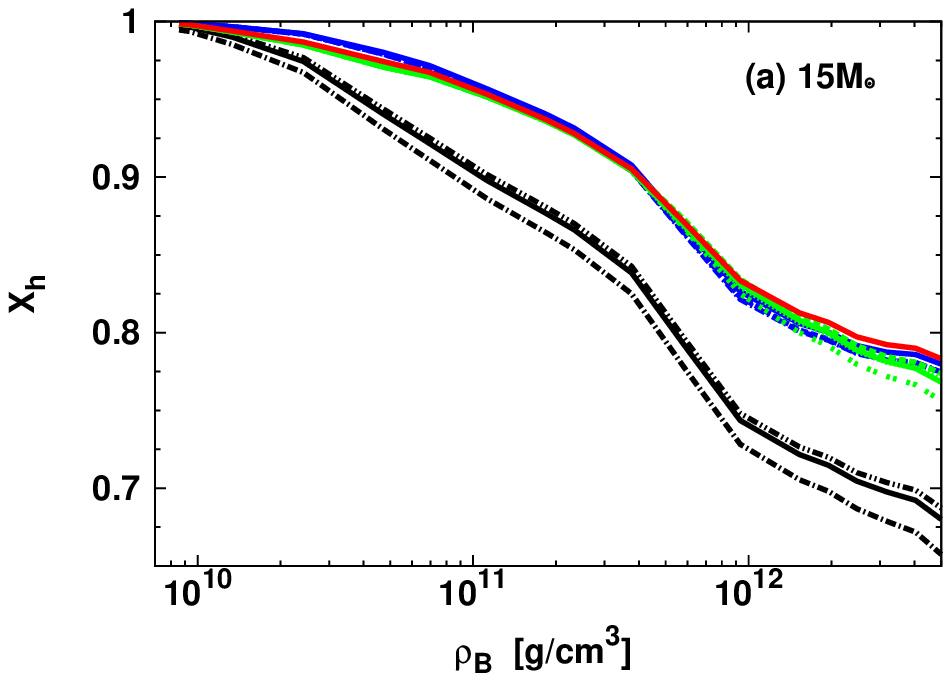}
\includegraphics[width=8cm]{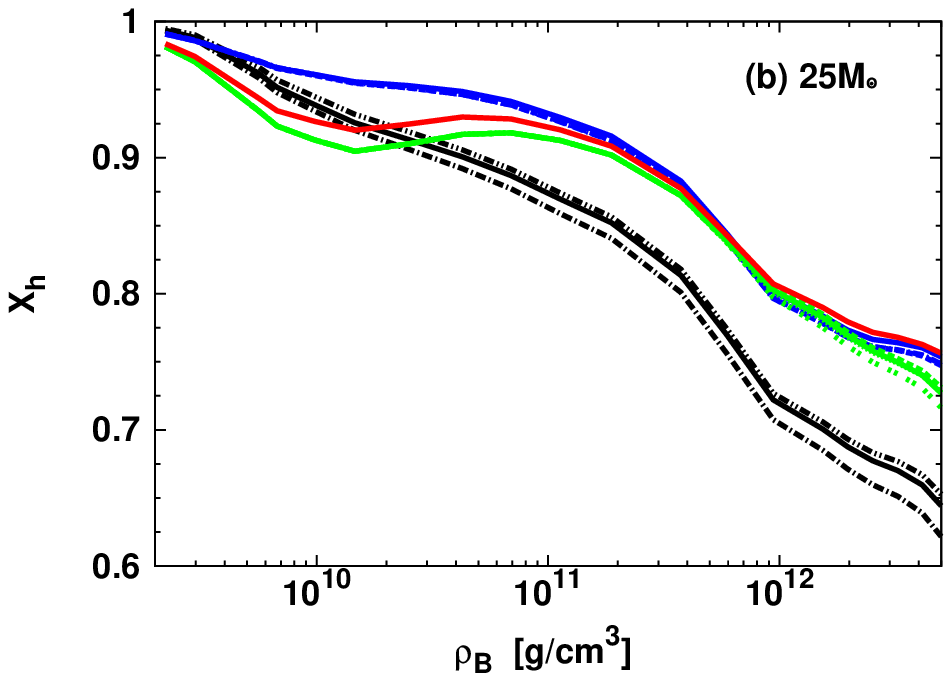}
%\ \\
%\ \\
\caption{Total mass fractions of heavy nuclei  at the centers of the collapsing cores of supernova progenitors of 15M$_\odot$ (left panel) and 25M$_\odot$ (right panel) for models  
1B (black dashed-dotted lines), 1D (black double-dotted dashed lines), 1E (black solid lines),
 2FB (blue dashed-dotted lines),  2FE (blue solid lines),  2KE (blue dashed lines), 
 3FE (green solid lines), 3KE (green dashed lines), 3HE (green dotted lines), 3WE (green double-dotted lines), and  4FE (red solid lines). }
\label{fig_xh}
\end{figure}

Figure~\ref{fig_xh} shows the total mass fraction of heavy nuclei, $X_h=\sum_{Z>5,N>5}X_{AZ}$.
It is found that the mass fraction is generally large for a model that has many
internal degrees of freedom.
The difference is greater for the 25M$_\odot$ star in which the central temperature  is higher than that in the 15M$_\odot$ star.
Model 4FE always reproduces larger mass fractions than model 3FE. % and 3FE  
At the beginning of core collapse, models 2FB, 2FE, and 2KE yield larger mass fractions than models 3FE or 4FE 
 because  of the larger internal degrees of freedom,
 while the mass fraction for model 4FE becomes larger than those of models 2 
 at $\rho_B\approx10^{12}{\rm g/cm}^3$,  because  nuclei with large mass numbers  are
 populated in model 4FE, as shown in Figs.~\ref{figlvl} and \ref{fig_dis15}.   
The mass fractions for models 1B, 1D, and 1E deviate from those for the other models,  because shell effects are not considered.
%}%%end
At low densities in models 2 and 3, the choice of mass data hardly affects the mass fraction
 because of the fact that the internal free energies of the nuclei are almost identical to
 their experimental values of nuclear masses and  the nuclei are not excited at low temperatures.
At $\rho_B\approx10^{12}{\rm g/cm}^3$, slight differences arising from the mass data appear.
%
%in models 1B, 1D, 1E, 2FB, and 2FE,  because of  them. %bulk property differences. 
%Although the difference are just in a few percent,
It is found that softer bulk properties increase the mass fractions of heavy nuclei, 
in the order of models 1D, 1E, and  1B  (2FD, 2FE, and 2FB), 
which corresponds to the ascending order of  $S_0$ or $L$. 
%or of densities at which the bulk energies of asymmetric matter increase rapidly, as shown in Fig~\ref{fig_blk}.
%The impact of nuclear mass data on the total mass fraction are much smaller than those of other ingredients even at high densities.
%chemical potentials ?
%pressure ?

\begin{figure}
\includegraphics[width=8cm]{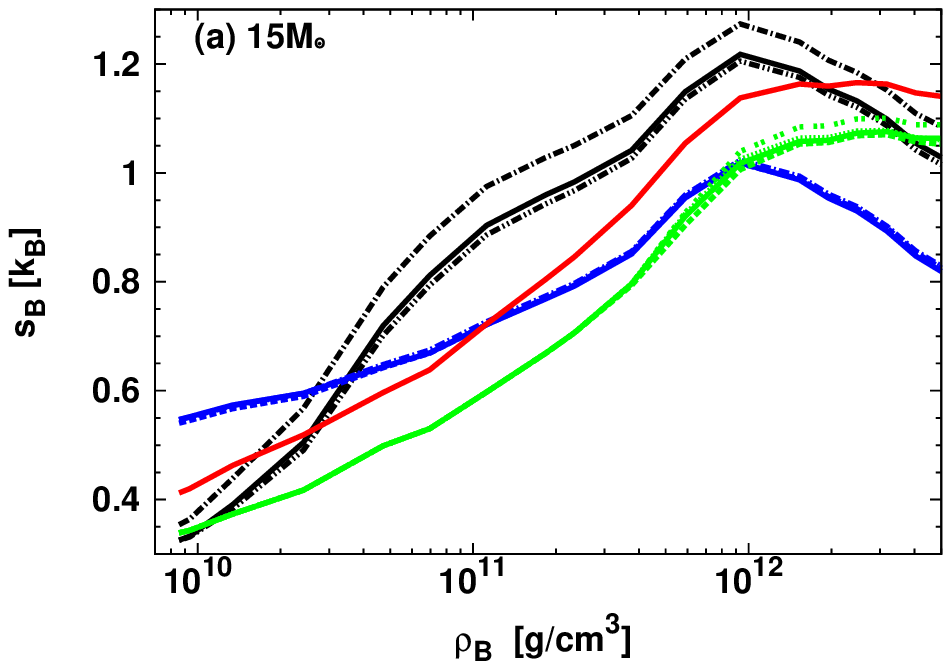}
\includegraphics[width=8cm]{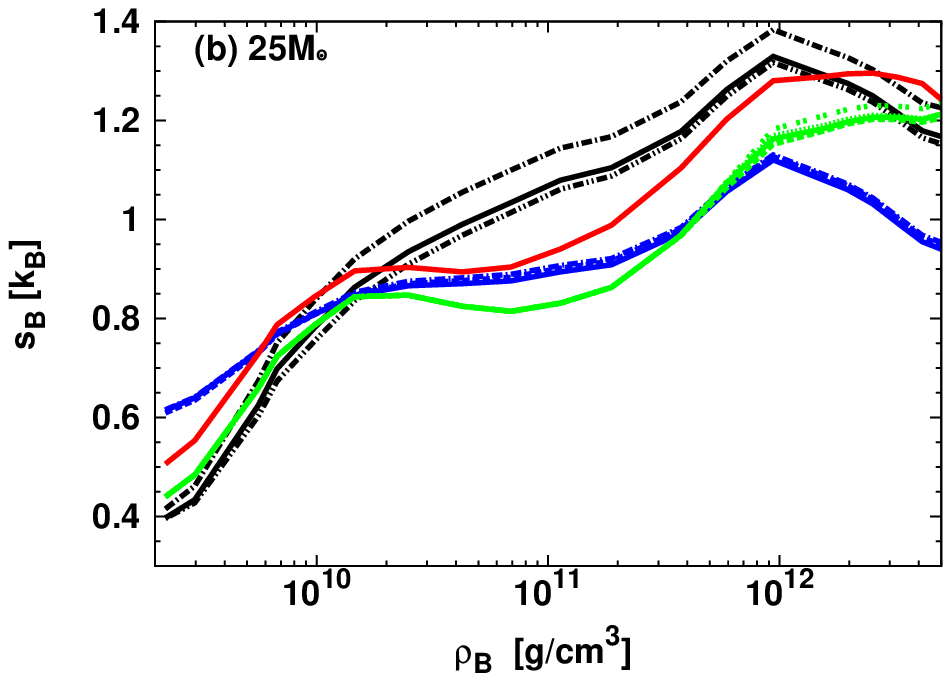}
%\ \\
\caption{Baryonic entropies per baryon at the centers of the collapsing cores of supernova progenitors of 15M$_\odot$ (left panel) and 25M$_\odot$ (right panel) for models  
1B (black dashed-dotted lines), 1D (black double-dotted dashed lines), 1E (black solid lines),
 2FB (blue dashed-dotted lines),  2FE (blue solid lines),  2KE (blue dashed lines), 
 3FE (green solid lines), 3KE (green dashed lines), 3HE (green dotted lines), 3WE (green double-dotted lines), and  4FE (red solid lines). }
\label{fig_ent}
\end{figure}

%In actual supernova simulations, 
The central densities and temperatures of  the collapsing cores increase adiabatically, 
and the entropy is an essential quantity for determining the dynamics of core collapse.
I therefore compare the baryonic entropies per baryon in Fig.~\ref{fig_ent}. 
These are given by $s_B=-\partial f/\partial T|_{n_B,Y_p}/n_B$ and are essentially determined by the kinetic terms of the nuclear components and the temperature derivatives of the internal degrees of freedom, $\partial g_{AZ}(T)/\partial T$, and  of  the internal free energy, $\partial F_{AZ}(T)/\partial T$.
Note from the figure that the entropies differ among the models based on 
different finite-temperature modeling, even in  the initial stages of core collapse.
For the models with smaller $X_h$, such as for models 1B, 1D, and 1E,
the entropy is more likely to be high 
because of  the increase in the population of dripped neutrons.
Models 2FB, 2FE, and 2FK at  at $\rho_B\approx10^{10}{\rm g/cm}^3$ and model 4FE at $\rho_B\approx10^{12}{\rm g/cm}^3$  tend to yield  high entropies  
because of the large values of  $\partial g^*_{AZ}(T)/\partial T$ and  $\partial g_{AZ}(T)/\partial T$.
 These differences in entropy would be  influential for the fate of core-collapse supernovae  \cite{suwa16, furusawa17d,nagakura18b}.

\section{Summary and Discussion \label{sec:conc}}
I have compared the 14 statistical EOSs for the core collapse of a massive star by systematically changing the bulk properties of nuclear matter, 
the masses of neutron-rich nuclei, and  the treatment of finite-temperature effects in the nuclear model.
Overall, the temperature dependencies of the internal degrees of freedom
 and of the internal free energy are paramount  to determining the entropy and nuclear composition during the core-collapse phase.
 Differences in these quantities among EOSs would affect thermodynamic conditions
 and ensemble-averaged weak-interaction rates of nuclear electron-captures and neutrino-nucleus coherent scattering \cite{furusawa17b};
 therefore, the dynamics of the core collapse would be sensitive to the treatment of the finite-temperature effects.
The parameters of bulk nuclear matter and the theoretical mass data  for neutron-rich nuclei do not greatly influence the average  mass  number or the total mass fraction of heavy nuclei. %, at least in the LDM EOS models. 
%The softness of nuclear bulk properties, however, causes a little change in the average mass number and a little increase in the total mass fraction of heavy nuclei.

Semi-empirical   expression for internal degrees of freedom that ignore the temperature dependence of shell effects may not be appropriate for discussions of core-collapse nuclei, 
because they overestimate  the number densities of magic nuclei.
More precise calculations for internal partition functions are required.
The individual level densities of medium-mass, neutron-rich nuclei  however,  have not
been well studied.
Actually, some important nuclei  lack the data \cite{rauscher03} 
used in the present work.
Phenomenological models with nuclear shell washout 
%internal free energies that depend on temperature
 tend to increase the populations of nuclei with small mass numbers that have 
large electron-capture rates; they would therefore reduce the charge fraction and entropy significantly in collapse simulations \cite{furusawa17b,nagakura18b}.
The washout formulation for the nuclear shell energies, however, is very simple and some parts, especially its mass-number dependence, should be improved.

Some studies based on EOSs with semi-empirical expression for internal degrees of freedom \cite{sullivan16,titus18} 
indicate that  nuclei above  the double-magic nuclei, $(N,Z)=(50,28)$ in the nuclear chart are primary targets for studies of electron-capture during core-collapse.
In all the EOSs discussed in this works, such nuclei  as $(N,Z)=(50,30)$ are certainly populated, while non-magic nuclei are also abundant
 in the EOSs with  more-reliable finite-temperature modeling.
% and in the EOSs based on the LDM with shell washout.
Non-magic nuclei, such as $(N,Z)=(40,25)$ and  $(60,35)$, may also contribute
to the deleptonization of the core  \cite{furusawa17b}.
I conclude that further investigations of the internal partition functions of nuclei with proton numbers 25--45 and the neutron numbers 40--85 at $T\approx$0.5--3.0 MeV are needed to remove one of the serious ambiguities in the input physics for core-collapse supernova studies.
Finite-temperature effects may also change the electron-capture rates themselves in addition to the nuclear abundances  \cite{langanke03,dzhioev10,furusawa17b,titus18}. 

It is surprising that  the choice of   finite-temperature modeling  produces significant
 differences in  the entropy and nuclear composition, even at the beginning of core  collapse.
 To perform  supernova simulations consistent with stellar evolution calculations, the same internal partition functions should be utilized in both stages.
Even in the postprocessing nucleosynthesis calculations for the ejecta of  supernova explosions, 
the thermodynamic conditions at low temperatures and densities should be shared with the dynamical simulations.
The setup of mass data and internal degrees of freedom also needs to  be better unified for  the stages immediately following  supernova explosions.

Note that this comparative study is incomplete to cover all nuclear uncertainties.
For instance, the bulk free energy of uniform nuclear matter depends not only 
on bulk properties at nuclear saturations but also on the theoretical approach such as relativistic mean-field theory, variational method, or chiral effective theory \cite{holt17}. 
In addition, there are many other works for the nuclear physics inputs, such as 
a comprehensive study of nuclear level densities based on two backshifted Fermi gas models and a constant-temperature model  \cite{egidy05}.
%{\bf
%The data would be helpful to calculate the internal partition functions in EOS for supernova matter.
 Very recently, Raduta and Gullemineli \cite{raduta18} have constructed EOS data for supernova simulations based on
 Gullemineli and Raduta \cite{gulminelli15} with the level-density data.  
%and 
Further comparison of the EOSs  with a focus on the level densities would be interesting. 
%}%\endbf
%}

In this work, I have  systematically compared several ingredients as independent inputs for EOS models. 
In real, the mass data, partition functions, and nuclear matter calculations
should  be related to each other.  
%Ideally, %we should perform comparisons of E
%the EOS should be based on a specific nuclear model that is consistent with mass data, partition functions, and nuclear matter calculations.  
For instance, Rauscher's partition function is calculated by using the nucleon separation energies in the old version of the FRDM mass data, and 
the WS4 mass data is based on a nuclear model with individual bulk properties, such as %e.g., 
$S_0$=30.16 MeV. 
Level-density data consistent with HFB mass data
are also provided, which are based on Hartree-Fock-Bogoliubov theory  \cite{goriely08}.
%
% in the nuclear model for WS4 mass data.
Ideally, comparisons of self-consistent EOSs based on a specific nuclear model
should be performed.
% may help to determine the nuclear properties  from future neutrino observations of core-collapse supernovae.
At present, however, first-principles calculations of heavy nuclei have not been completed, even for ground-state nuclei.
More realistic supernova simulations, and  predictions that can be compared with  neutrino and gravitational wave observations,  will require
%we should  improve
step-by-step improvements in the calculations of nuclear matter, nuclear masses, internal degrees of freedom and weak-interaction rates as nuclear physics inputs.
% step by step.

\begin{acknowledgments}
 The author acknowledges M. Hempel, I. N. Mishustin, K. Sumiyoshi, S. Yamada, H. Suzuki, S. Typel, G. Martinez-Pinedo, H. Sotani, Y. Yamamoto, H. Nagakura, H. Togashi, Z. Niu, and C. Kato for fruitful discussion. 
The author is grateful to the anonymous referee for his or her careful
reading of the manuscript and helpful comments.
This work was supported by JSPS KAKENHI (Grant No. JP17H06365)
and  HPCI Strategic Program of Japanese MEXT  (Project ID: hp170304, 180111).
A part of the numerical calculations was carried out on  PC cluster at Center
for Computational Astrophysics, National Astronomical Observatory of Japan and XC40 at YITP in Kyoto University.
\end{acknowledgments}

\bibliography{reference180926}

\newpage

\begin{table}[t]
\begin{tabular}{|c||c|c|c||c|c|}
\hline 
\hline
  Model & finite-temperature modeling  & mass data  &  bulk parameter &
 $\sigma_0$ [MeV/fm$^{2}$] 
& $C_s$\\
 \hline
1B & SMSM & --  & B & -- & --  \\
1D & SMSM &  -- & D  &--  & --  \\
1E & SMSM &  -- & E  & --  & --  \\
2FB & LDM + washout &  FRDM &  B & 1.047 &  19.66 \\
2FD & LDM + washout  & FRDM  & D & 1.052 & 29.68  \\
2FE & LDM + washout &  FRDM & E  & 1.042 &  28.36 \\
2HE & LDM + washout & HFB24 & E  & 1.042 &  27.12 \\
2KE  & LDM + washout & KTUY & E & 1.042 &  32.16 \\
2WE & LDM + washout &  WS4 &  E  & 1.042 & 28.91 \\
3FE &Fai $\&$ Randrup &  FRDM & E & --  & --   \\
3HE &Fai $\&$ Randrup  & HFB24 & E & --  & --   \\
3KE &Fai $\&$ Randrup  & KTUY & E & --  & --   \\
3WE  &Fai $\&$ Randrup & WS4 & E & --  & --   \\
4FE  & Rauscher & FRDM & E & -- & -- \\
 \hline
%\hline 
\hline
\end{tabular}
\caption{\label{tab_model}%
List of models used for systematic comparisons.
The first four columns %of $n_{eqi}$, $T_c$,  $\gamma(T)$, and light clusters
provide the model name, the treatment of finite-temperature effects (the SMSM EOS \cite{botvina10}, a LDM with shell-washout \cite{furusawa17a}, a semi-empirical formula by Fai and Randrup  \cite{fai82}, or a Fermi-gas approach by Rauscher \cite{rauscher03}),  the theoretical mass data (FRDM \cite{moller16}, HFB24 \cite{goriely13}, KTUY \cite{koura05}, or WS4 \cite{wang14}), and 
the parameter set used for bulk nuclear matter  (B, D, or E \cite{oyamatsu03,oyamatsu07}).
The last two columns provide
the values of the surface tension $\sigma_0$ for symmetric nuclei and the isospin-dependence parameter $C_s$, both for model 2.} 
%The other parameter is set to be $b_3=$1.58632 fm$^3$.}  
\end{table}

\begin{table}[t]
\begin{tabular}{|c||c|c|c|}
\hline 
\hline
 parameter set  & B & D &E  \\
 \hline
  $n_{0}$ [fm$^{-3}$]  &  0.15969 & 0.16905 & 0.15979 \\
 $\omega_0$ [MeV] &-16.184  & -16.224 &   -16.145 \\
 $K_0$ [MeV] & 230 & 180  &  230  \\
 $S_0$ [MeV]  & 33.550 & 30.543 &  31.002  \\
 $L$ [MeV] & 73.214 & 30.974   & 42.498 \\
 $-y$ [MeV fm$^3$] & 220 & 350 & 350 \\
 \hline
%\hline 
\hline
\end{tabular}
\caption{\label{tab_bulk}%
Parameters of bulk nuclear matter \cite{oyamatsu03,oyamatsu07}.} 
%The other parameter is set to be $b_3=$1.58632 fm$^3$.}  
\end{table}

\end{document}